\documentstyle[12pt]{article}
\begin{document}

\begin{titlepage}

\vspace*{-2cm}

\hspace*{\fill} CGPG-96/1-7\\
\hspace*{\fill} UNC-MATH-97/3
\vspace{.5cm}

\begin{centering}

{\huge The sh Lie structure of Poisson brackets in field theory}  

\vspace{.5cm}
{\large G. Barnich$^{1,*}$, R. Fulp$^2$, T. Lada$^2$ and
  J. Stasheff$^{3,\dagger}$}\\  

\vspace{.5cm}

$^1$ Center for Gravitational Physics and Geometry, The Pennsylvania
State University, 104 Davey Laboratory, University Park, PA 16802. \\ 

\vspace{.5cm} 

$^2$ Department of Mathematics, North Carolina State University,
Raleigh, NC 27695-8205.\\  

\vspace{.5cm}

$^3$ Department of Mathematics, The University of North Carolina at   
Chapel Hill, Phillips Hall, Chapel Hill, NC 27599-3250.

\vspace{.5cm}

\begin{abstract}
A general construction of an sh Lie algebra ($L_{\infty}$-algebra) from 
a homological resolution of a Lie algebra is given. It is
applied to the space of local functionals
equipped with a Poisson bracket, induced by a bracket for
local functions along the lines suggested by Gel'fand, Dickey and
Dorfman. In this way, higher order maps are constructed which combine
to form an sh Lie algebra on the graded differential algebra of
horizontal forms. The same construction applies for graded brackets in
field theory such as the Batalin-Fradkin-Vilkovisky bracket of the
Hamiltonian BRST theory or the Batalin-Vilkovisky antibracket.
\end{abstract}

\end{centering}

\vspace{1.5cm}

{\footnotesize \hspace{-0.6cm}($^*$) Research supported by
  grants from the Fonds National Belge de la Recherche Scientifique
  and the Alexander-von-Humboldt foundation. New address: Freie
  Universit\"at Berlin, Institut f\"ur Theoretische Physik, Arnimallee
  14, D-14195 Berlin.}  

{\footnotesize \hspace{-0.6cm}($^\dagger$) Research supported in part
  by NSF grant DMS-9504871.}

\end{titlepage}

\pagebreak

\def\lh{\hbox to 15pt{\vbox{\vskip 6pt\hrule width 6.5pt height 1pt}
    \kern -4.0pt\vrule height 8pt width 1pt\hfil}}
    \def\blob{\mbox{$\;\Box$}} 
\def\qed{\hbox{${\vcenter{\vbox{\hrule height 0.4pt\hbox{\vrule width
    0.4pt height 6pt \kern5pt\vrule width 0.4pt}\hrule height
    0.4pt}}}$}} 
\newtheorem{theorem}{Theorem}
\newtheorem{lemma}[theorem]{Lemma}
\newtheorem{definition}[theorem]{Definition}
\newtheorem{corollary}[theorem]{Corollary} 
\newcommand{\proof}[1]{{\bf Proof:} #1~$\qed$}
    \newtheorem{proposition}[theorem]{Proposition} \def\CJ{Loc(E)} 

\section{Introduction}

In field theories, an important class of physically interesting
quantities, such as the action or the Hamiltonian, are described by
local functionals, which are the integral over some region of
spacetime (or just of space) of local functions, i.e., functions which
depend on the fields and a finite number of their derivatives. It is
often more convenient to work with these integrands instead of the
functionals, because they live on finite dimensional spaces. The price
to pay is that one has to consider equivalence classes of such
integrands modulo total divergences in order to have a one-to-one
correspondence with the local functionals. 

The approach to Poisson brackets in this context, pioneered by
Gel'fand, Dickey and Dorfman, is to consider the Poisson brackets for
local functionals as being induced by brackets for local functions,
which are not necessarily strictly Poisson. We will analyze here in
detail the structure of the brackets for local functions corresponding
to the Poisson brackets for local functionals. More precisely, we will
show that these brackets will imply higher order brackets combining
into a strong homotopy Lie algebra. 

The paper is organized as follows:

In the case of a homological resolution of a Lie algebra, it is shown
that a skew-symmetric bilinear map on the resolution space inducing
the Lie bracket of the algebra extends to higher order
`multi-brackets' on the resolution space which combine to form an
$L_\infty$-algebra (strong homotopy Lie algebra or sh Lie
algebra). For completeness, the definition of these algebras is
briefly recalled. For a highly connected complex which is not a
resolution, the same procedure yields part of an sh Lie algebra on
the complex with corresponding multi-brackets on the homology .

This general construction is then applied in the following case.

If the horizontal complex of the variational bicomplex is used as
a resolution for local functionals equipped with a Poisson bracket, we
can construct an sh Lie algebra (of order the dimension of the base
space plus two) on the graded differential algebra of horizontal
forms. The construction applies not only for brackets in Darboux
coordinates as well as for non-canonical brackets such as those of the KDV 
equation, but also in the presence of Grassmann odd
fields for graded even brackets, such as the extended Poisson bracket
appearing in the Hamiltonian formulation of the BRST theory, and
for graded odd brackets, such as the antibracket of the
Batalin-Vilkovisky formalism. 

\section{Sh Lie algebras from homological resolution of Lie algebras}

\subsection{Construction}

Let ${\cal F}$ be a vector space and $(X_*,l_1)$ a homological
resolution thereof, i.e., $X_*$ is a graded vector space, $l_1$ is a 
differential and lowers the grading by one with ${\cal F}\simeq
H_0(l_1)$ and $H_k(l_1)=0$ for $k>0$. The complex $(X_*,l_1)$ is
called the resolution space.  (We are {\it not} using the term `resolution'
in a categorical sense.)

Let ${\cal C}_*$ and ${\cal B}_*$ denote the $l_1$ cycles
(respectively, boundaries) of $X_*$. Recall that by convention $X_0$
consists entirely of cycles, equivalently $X_{-1}=0$. Hence, we have a
decomposition 
\begin{eqnarray} X_0={\cal B}_0\oplus{\cal K},
\end{eqnarray}
with ${\cal K}\simeq {\cal F}$.

We may rephrase the above situation in terms of the existence of a
contracting homotopy on $(X_*,l_1)$ specifying a homotopy inverse for
the canonical homomorphism $\eta:X_0\longrightarrow H_0(X_*)\simeq
{\cal F}$. We may regard ${\cal F}$ as a differential graded vector
space ${\cal F}_*$ with ${\cal F}_0={\cal F}$ and ${\cal F}_k=0$ for
$k>0$; the differential is given by the trivial map. We then consider
the chain map $\eta:X_*\longrightarrow {\cal F}_*$ with homotopy
inverse $\lambda:{\cal F}_*\longrightarrow X_*$; i.e., we have that
$\eta \circ \lambda = 1_{{\cal F}_*}$ and that $\lambda \circ \eta
\sim 1_{X_*}$ via a chain homotopy $s:X_*\longrightarrow X_*$ with
$\lambda \circ \eta -1_{X_*}=l_1\circ s +s \circ l_1$.
Observe that this equation takes on the form $\lambda \circ \eta
-1_{X_*}=l_1 \circ s$ on $X_0$.

We may summarize all of the above with the commutative diagram

$$
\begin{array}{cccccc}
&&s& &s\\
\cdots\longrightarrow & X_2 &\stackrel
{\textstyle\longleftarrow}{\longrightarrow} &X_1&\stackrel
{\textstyle\longleftarrow}{\longrightarrow} &X_0\\ & & l_1 & & l_1&\\
& \lambda\Bigl\uparrow\Bigr\downarrow\eta & &
\lambda\Bigl\uparrow\Bigr \downarrow\eta&&
\lambda\Bigl\uparrow\Bigr\downarrow\eta\\ &&&&&\\
\cdots\longrightarrow &0&\longrightarrow & 0 &\longrightarrow &
H_0={\cal F}. 
\end{array}
$$

It is clear that $\eta (b)=0$ for $b \in {\cal B}_0$.

Let $(-1)^{\sigma}$ be the signature of a permutation $\sigma$ and
$unsh(k,p)$ the set of permutations $\sigma$ satisfying
$$
\underbrace {\sigma(1)<\dots<\sigma(k)}_{\rm first\ \sigma\ hand}
\hspace{1cm}{\rm and}\hspace{1cm} \underbrace
{\sigma(k+1)<\dots<\sigma(k+p).}_{\rm second\ \sigma\ hand} 
$$ 

We will be concerned with skew-symmetric linear maps 
\begin{eqnarray} 
\tilde l_2:X_0 \otimes X_0\longrightarrow X_0 
\end{eqnarray} that satisfy the properties
\begin{eqnarray}
& & (i)\ \tilde l_2(c,b_1)=b_2\label{y}\\ & & (ii)\ \sum_{\sigma \in 
unsh(2,1)}(-1)^{\sigma} \tilde l_2(\tilde l_2(c_{\sigma(1)}, 
c_{\sigma(2)}),c_{\sigma(3)})=b_3 \label{a} 
\end{eqnarray} 
where $c, c_i\in X_0$, $b_i\in {\cal B}_0$ and with the additional
structures on $X_*$ as well as on ${\cal F}$ that such maps will yield. 

We begin with
\begin{lemma}
The existence of a skew-symmetric linear map $\tilde l_2:X_0\bigotimes
X_0 \longrightarrow X_0$ that satisfies condition $(i)$ is equivalent
to the existence of a skew-symmetric linear map 
$ [\cdot,\cdot]:{\cal F}\bigotimes {\cal F}\longrightarrow {\cal F}$. 
\end{lemma}
\proof{$\tilde l_2$ induces a linear mapping on ${\cal F}\bigotimes
  {\cal F}$ via the diagram
$$
\begin{array}{ccc}
X_0\bigotimes X_0 & \stackrel {\tilde l_2}{\longrightarrow}& X_0\\
\lambda \otimes \lambda \uparrow& & \downarrow \eta\\ {\cal
  F}\bigotimes {\cal F} & \stackrel {[\cdot,\cdot]}
{\longrightarrow}&{\cal F.} \end{array} $$ 

The fact that $\tilde l_2$ satisfies condition $(i)$ guarantees that
$[\cdot,\cdot]$ is well-defined on the homology classes.

Conversely, given $[\cdot,\cdot]:{\cal F}\bigotimes {\cal
  F}\longrightarrow {\cal F}$, define $\tilde l_2=\lambda \circ
  [\cdot,\cdot]\circ \eta \otimes \eta$. It is clear that $\tilde l_2$
  is skew-symmetric because $[\cdot,\cdot] $ is, and condition $(i)$
  is satisfied in the strong sense that $\tilde l_2(c,b)=0$.} 

\begin{lemma}
Assume that $\tilde l_2:X_0\bigotimes X_0 \longrightarrow X_0$
satisfies condition $(i)$. Then the induced bracket on ${\cal F}$ is a
Lie bracket if and only if $\tilde l_2$ satisfies condition $(ii)$.
\end{lemma}
\proof{Assume that the induced bracket on ${\cal F}$ is a Lie bracket;
  recall that the bracket is given by the composition $\eta \circ
  \tilde l_2\circ \lambda\otimes\lambda$. The Jacobi identity takes on
  the form, for arbitrary $f_i\in {\cal F},$ 
\begin{eqnarray}
\sum_{\sigma\in unsh(2,1)}(-1)^{\sigma}(\eta \circ \tilde l_2\circ
\lambda\otimes\lambda) (\eta \circ (\tilde l_2\otimes 1)\circ
(\lambda\otimes\lambda\otimes 1)
\nonumber\\(f_{\sigma(1)}\otimes f_{\sigma(2)}\otimes
f_{\sigma(3)})=0\nonumber\\ \Leftrightarrow \sum_{\sigma\in
unsh(2,1)}(-1)^{\sigma}(\eta \circ \tilde l_2\circ \lambda\otimes\lambda)
(\eta \circ \tilde l_2( \lambda (f_{\sigma(1)}) \otimes \lambda
(f_{\sigma(2)})) \otimes f_{\sigma(3)})=0\nonumber\\ \Leftrightarrow
\sum_{\sigma\in unsh(2,1)}(-1)^{\sigma}\eta \circ \tilde l_2
(\lambda\circ \eta\circ \tilde l_2(\lambda (f_{\sigma(1)}) \otimes
\lambda (f_{\sigma(2)}))\otimes \lambda (f_{\sigma(3)}) )
=0\nonumber\\ \Leftrightarrow \sum_{\sigma\in
unsh(2,1)}(-1)^{\sigma}\eta\circ \tilde \l_2((1+l_1\circ s) \circ
\tilde l_2 (\lambda (f_{\sigma(1)})\otimes\lambda
(f_{\sigma(2)}))\otimes\lambda( f_{\sigma(3)}))=0\nonumber\\
\Leftrightarrow \sum_{\sigma\in unsh(2,1)}(-1)^{\sigma}\eta
\circ\tilde \l_2( \tilde
l_2(\lambda(f_{\sigma(1)})\otimes\lambda(f_{\sigma(2)}))\otimes\lambda(
f_{\sigma(3)}))\nonumber\\ +\sum_{\sigma\in
  unsh(2,1)}(-1)^{\sigma}\eta\circ \tilde \l_2(l_1\circ s\circ \tilde
l_2(\lambda(f_{\sigma(1)})\otimes\lambda(f_{\sigma(2)}))\otimes\lambda(
f_{\sigma(3)}))=0\nonumber\\ \Leftrightarrow \eta(\sum_{\sigma\in
unsh(2,1)}(-1)^{\sigma}\tilde \l_2( \tilde
l_2(\lambda(f_{\sigma(1)})\otimes\lambda(f_{\sigma(2)}))\otimes\lambda(
f_{\sigma(3)})))+\eta (b)=0\nonumber
\end{eqnarray}
But $\eta (b)=0$ and so
\begin{eqnarray}
\sum_{\sigma\in unsh(2,1)}(-1)^{\sigma}\tilde \l_2( \tilde
l_2(\lambda(f_{\sigma(1)})\otimes\lambda(f_{\sigma(2)}))\otimes\lambda(
f_{\sigma(3)}))=b^\prime \in {\cal B}_0. 
\end{eqnarray} 
The converse follows from a similar calculation.}

${\bf Remark:}$
The interesting case here is when ${\cal F}$ is only known as
$X_0/{\cal B}_0$ and the only characterization of the Lie bracket
$[\cdot,\cdot]$ in ${\cal F}$ is as the bracket induced by $\tilde
l_2$. An important particular case, to be considered elsewhere,
occurs when $X_0$ is a Lie algebra ${\cal G}$ with Lie
bracket $L_2$ and ${\cal B}_0$ a Lie ideal. The bracket $\tilde l_2$
is defined by choosing a vector space complement ${\cal K}$ of the
ideal ${\cal B}_0$ in $ X_0$ and then projecting the Lie bracket
$L_2$ onto ${\cal K}$. Hence, $L_2(c_1,c_2)=\tilde
l_2(c_1,c_2)+b(c_1,c_2)$, where $b(c_1,c_2)$ is a well-defined element
of ${\cal B}_0$. Indeed, by definition, property (i) holds with zero
on the right hand side. Property (ii) follows from the
Jacobi identity for $L_2$~: 
\begin{eqnarray}
0=\sum_{\sigma \in unsh(2,1)}(-1)^{\sigma} L_2(L_2(c_{\sigma(1)},
c_{\sigma(2)}),c_{\sigma(3)})\nonumber\\=\sum_{\sigma \in
unsh(2,1)}(-1)^{\sigma} [\tilde l_2(L_2(c_{\sigma(1)},
c_{\sigma(2)}),c_{\sigma(3)})+b(L_2(c_{\sigma(1)},
c_{\sigma(2)}),c_{\sigma(3)})]\nonumber\\=\sum_{\sigma \in
unsh(2,1)}(-1)^{\sigma} [\tilde l_2(\tilde l_2(c_{\sigma(1)},
c_{\sigma(2)}),c_{\sigma(3)})+b(L_2(c_{\sigma(1)},
c_{\sigma(2)}),c_{\sigma(3)})]. 
\end{eqnarray}

We now turn our attention to the maps that $\tilde l_2$ induces on the
complex $X_*$.
\begin{lemma}
A skew-symmetric linear map $\tilde l_2:X_0\bigotimes
X_0\longrightarrow X_0$ that satisfies condition $(i)$ extends to a
degree zero skew-symmetric chain map $l_2:X_*\bigotimes X_*
\longrightarrow X_*$. 
\end{lemma}
\proof{We first extend $\tilde l_2$ to a linear map $l_2:X_1\bigotimes
X_0  \longrightarrow X_1$ by the following: let $x_1\otimes x_0 \in
X_1\bigotimes X_0$. Then $l_1(x_1\otimes x_0)=l_1x_1\otimes
x_0+x_1\otimes l_1x_0=l_1x_1\otimes x_0\in X_0\bigotimes X_0$.
So we have that $l_2l_1(x_1\otimes x_0)=\tilde l_2(l_1x_1\otimes
x_0)=b$ by condition $(i)$. Write $b=l_1z_1$ for $z_1\in X_1$ and
define $l_2(x_1\otimes x_0)=z_1$. Also extend $l_2$ to $X_0\bigotimes
X_1$ by skew-symmetry: $l_2(x_0\otimes x_1)=-l_2(x_1\otimes
x_0)$. Note that $l_2$ is a chain map by construction. 

Now assume that $l_2$ is defined and is a chain map on elements of
degree less than or equal to $n$ in $X_*\bigotimes X_*$. Let
$x_p\otimes x_q\in X_p\bigotimes X_q$ where $p+q=n+1$. Because
$l_1(x_p\otimes x_q)$ has degree $n$, $l_2l_1(x_p\otimes x_q)$ is
defined. We have that 
\begin{eqnarray}
&&l_1l_2l_1(x_p\otimes x_q) = \nonumber\\
& &\ \ \ \ =l_1l_2[l_1x_p\otimes x_q+(-1)^px_p\otimes l_1x_q]
\nonumber\\ &  &\ \ \ \ =l_2l_1[l_1x_p\otimes x_q+(-1)^px_p\otimes
l_1x_q] {\rm \ \ \ because\ } l_2 {\rm \ is \ a\ chain\ map\
  }\nonumber\\ & &\ \ \ \ =l_2[l_1l_1x_p\otimes
x_q+(-1)^{p-1}l_1x_p\otimes l_1x_q\nonumber\\ & &\ \ \ \ \ \ \ \
+(-1)^pl_1x_p\otimes l_1x_q +x_p\otimes l_1l_1x_q]=0 
\end{eqnarray}
because $l_1^2=0$ and $(-1)^p$ and $(-1)^{p-1}$ have opposite
parity. Thus  $l_2l_1(x_p\otimes x_q)$ is a cycle in $X_n$ and so
there is an element $z_{n+1}\in X_{n+1}$ with
$l_1z_{n+1}=l_2l_1(x_p\otimes x_q)$. Define $l_2(x_p\otimes
x_q)=z_{n+1}$. As before, extend $l_2$ to $X_q\bigotimes X_p$ by
skew-symmetry and note that $l_2$ is a chain map by construction.} 

${\bf Remark :}$ We will be concerned with (graded) skew-symmetric
maps $f_n:\bigotimes^nX_* \longrightarrow X_*$ that have been extended to maps
$f_n:\bigotimes^{n+k}X_*\longrightarrow \bigotimes^kX_*$ via the
equation 
\begin{eqnarray}
& & f_n(x_1\otimes \dots \otimes x_{n+k})=\nonumber\\ &
&\sum_{unsh{(n,k)}}(-1)^{\sigma}e(\sigma) f_n(x_{\sigma(1)}
\otimes\dots\otimes x_{\sigma(n)})\otimes
x_{\sigma(n+1)}\otimes\dots\otimes x_{\sigma(n+k)}, 
\end{eqnarray} 
where $e(\sigma)$ is the Koszul sign  (see
e.g. \cite{lada-markl}). This extension arises from the
skew-symmetrization of the extension of a linear map as a skew coderivation
on the tensor coalgebra on the graded vector space $X_*$
\cite{lars}. The extension of the differential $l_1$ assumed in the
previous lemma is equivalent to the one given by this construction. We
assume for the remainder of this section that all maps have been
extended in this fashion when necessary; moreover, we will use the
uniqueness of such extensions when needed. 

When the original skew-symmetric map $\tilde l_2$ satisfies both
conditions $(i)$ and $(ii)$, there is a very rich algebraic structure
on the complex $X_*$. 
\begin{proposition}
A skew-symmetric linear map $\tilde l_2:X_0\bigotimes
X_0\longrightarrow X_0$ that satisfies conditions $(i)$ and $(ii)$
extends to a chain map $l_2:X_*\bigotimes X_* \longrightarrow X_*$;
moreover, there exists a degree one map $l_3: X_*\bigotimes X_*
\bigotimes X_*\longrightarrow X_*$ with the property that
$l_1l_3+l_3l_1+l_2l_2=0$. 
\end{proposition}

Here, we have suppressed the notation that is used to indicate the
indexing of the summands over the appropriate unshuffles as well as
the corresponding signs. They are given explicitly in Definition 5
below. 

\proof{We extend $\tilde l_2$ to $l_2:X_*\bigotimes X_*\longrightarrow
X_*$ as in the previous lemma. In degree zero, $l_2l_2(x_1\otimes
x_2\otimes x_3)$ is equal to a boundary $b$ in $X_0$ by condition
$(ii)$. There exists an element $z\in X_1$ with $l_1z=b$ and so we
define $l_3(x_1\otimes x_2\otimes x_3)=-z$. Because $l_1=0$ on
$X_0\bigotimes X_0\bigotimes X_0$, we have that
$l_1l_3+l_2l_2+l_3l_1=0$. 
 
Now assume that $l_3$ is defined up to degree $p$ in $X_*\bigotimes
X_*\bigotimes X_*$ and satisfies the relation
$l_1l_3+l_2l_2+l_3l_1=0$. Consider the map $l_2l_2+l_3l_1$ which is 
inductively defined on elements of degree $p+1$ in $X_*\bigotimes
X_*\bigotimes X_*$. We have that
$l_1[l_2l_2+l_3l_1]=l_1l_2l_2+l_1l_3l_1= 
l_2l_1l_2+l_1l_3l_1=l_2l_2l_1+l_1l_3l_1=
[l_2l_2+l_1l_3]l_1=-l_3l_1l_1=0$. Thus the image of $l_2l_2+l_3l_1$ is
a boundary in $X_p$ which is then the image of an element, say
$z_{p+1}\in X_{p+1}$. Define now $l_3$ applied to the original element
of degree $p+1$ in $X_*\bigotimes X_*\bigotimes X_*$ to be this
element $z_{p+1}$. } 

In the proof of the proposition above, we made repeated use of the
relation $l_1l_2-l_2l_1=0$ when extended to an arbitrary number of
variables. We may justify this by observing that the map
$l_1l_2-l_2l_1$ is the commutator of the skew coderivations $ l_1$ and
$l_2$ and is thus a coderivation; it follows that the extension of
this map must equal the extension of the $0$ map.

The relations among the maps $l_i$ that were generated in the previous
results are the first relations that one encounters in an sh Lie
algebra ($L_{\infty}$ algebra). Let us recall the definition
\cite{ls,lada-markl}. 
\begin{definition}
An sh Lie structure on a graded vector space $X_*$ is a collection of
linear, skew symmetric maps $l_k:\bigotimes ^kX_*\longrightarrow X_*$
of degree $k-2$ that satisfy the relation
\begin{eqnarray}
& & \sum_{i+j=n+1}\sum_{unsh{(i,n-i)}}
e(\sigma)(-1)^{\sigma}(-1)^{i(j-1)}
l_j(l_i(x_{\sigma(1)},\dots,
x_{\sigma(i)}),\dots ,x_{\sigma(n)}) =0,\nonumber \\ 
\end{eqnarray} where
$1\leq i,j$.
\end{definition}

{\bf Remark:} Recall that the suspension of a graded vector space
$X_*$, denoted by $\uparrow X_*$, is the graded vector space defined
by $(\uparrow X_*)_n=X_{n-1}$ while the desuspension of $X_*$ is given
by $(\downarrow X_*)_n=X_{n+1}$. It can be shown,
\cite{ls,lada-markl},  Theorem 2.3, that the data in the definition is
equivalent to 

a) the existence of a degree $-1$ coderivation $D$ on
$\bigwedge^*\uparrow X$, the cocommutative coalgebra on the graded
vector space $\uparrow X$, with $D^2=0$.

and to

b) the existence of a degree $+1$ derivation $\delta$ on
$\bigwedge^*\uparrow X^*$, the exterior algebra on the suspension
of the degree-wise dual of $X_*$, with $\delta^2=0$.  In this case, we require
that $X_*$ be of finite type. 

Let us examine the relations in the above definition independently of
the underlying vector space $X_*$ and write them in the form
$$\sum_{i+j=n+1}(-1)^{i(j-1)}l_il_j=0$$ where we are assuming that the
sums over the appropriate unshuffles with the corresponding signs are
incorporated into the definition of the extended maps $l_k$. We will
require the fact that the map
$$\sum_{i,j>1}(-1)^{i(j-1)}l_jl_i:\bigotimes^n X_*\longrightarrow
X_*$$ is a chain map in the following sense:

\begin{lemma}
Let $\{l_k\}$ be an sh Lie structure on the graded vector space
$X_*$. Then 
\begin{eqnarray}
l_1\sum_{i,j>1}(-1)^{(j-1)i}l_jl_i=(-1)^{n-1}
(\sum_{i,j>1}(-1)^{(j-1)i}l_j l_i)l_1
\end{eqnarray}
where $i+j=n+1$.
\end{lemma}
\proof{Let us reindex the left hand side of the above equation and
  write it as $$l_1\sum_{i=2}^{n-1}(-1)^{(n-i)i}l_{n-i+1}l_i$$ which,
  after applying the sh Lie relation to the composition
  $l_1l_{n-i+1}$, is equal to
  $$\sum_{i=2}^{n-1}\sum_{k=2}^{n-i}
  (-1)^{(n-i)i}(-1)^{(k-1)(n-i-k)+1}l_kl_{ n-i-k+2}l_i.$$ On the other
  hand, the right hand side may be written as
  $$(-1)^{n-1}\sum_{i=2}^{n-1}(-1)^{(i-1)(n-i+1)}l_il_{n-i+1}l_1$$
  which in turn is equal to  $$(-1)^{n-1}\sum_{i=2}^{n-1}
  \sum_{k=2}^{n-i}(-1)^{(i-1)(n-i+1)}
  (-1)^{(n-i-k+1)k}(-1)^{n-i+1}l_il_{n-i+2-k}l_k.$$ It is clear that
  the two resulting expressions have identical summands and a
  straightforward calculation yields that the signs as well are
  identical.} 

The argument in the previous proposition may be extended to construct 
higher order maps $l_k$ so that we have

\begin{theorem}
A skew-symmetric linear map $\tilde l_2:X_0\bigotimes
X_0\longrightarrow X_0$ that satisfies conditions $(i)$ and $(ii)$
extends to an sh Lie structure on the resolution space $X_*$.
\end{theorem}
\proof{We already have the required maps $l_1, l_2$ and $l_3$ from our
  previous work. We use induction to assume that we have the maps
  $l_k$ for $1\leq k< n$ that satisfy the relation in the definition
  of an sh Lie structure. To construct the map $l_n$, we begin with
  the map
  $$\sum_{i+j=n+1}(-1)^{(j-1)i}l_jl_i:\bigotimes^nX_0\longrightarrow
  X_{n-3}$$ with $i,j>1$. Apply the differential $l_1$ to this map to
  get  
\begin{eqnarray}
l_1\sum_{i+j=n+1}(-1)^{(j-1)i}l_jl_i=(-1)^{n-1}
\sum_{i+j=n+1}(-1)^{(j-1)i}l_jl_il_1=0
\end{eqnarray}
where the first equality follows from the lemma and the second
equality from the fact that $l_1$ is $0$ on $ \bigotimes^nX_0$. The
acyclicity of the complex $X_*$ will then yield, with care to preserve
the desired symmetry, the map $l_n$ on $\bigotimes^nX_*$ and it will
satisfy the sh Lie relations by construction.

Finally, assume that all of the maps $l_k$ for $k<n$ have been
constructed so as to satisfy the sh Lie relations and that $l_n$ has
been constructed in a similar fashion through degree $p$ in
$\bigotimes^nX_*$. We have the map
$$\sum_{i+j=n+1}(-1)^{(j-1)i}l_jl_i:(\bigotimes^nX_*)_p\longrightarrow
X_{p-3}$$ to which we may apply the differential $l_1$. This results in
$$\sum_{i+j=n+1}(-1)^{(j-1)i}l_1l_jl_i
=\sum_{i,j>1}(-1)^{(j-1)i}l_1l_jl_i+(-1)^{n-1}l_1l_nl_1$$ 
$$=\sum_{i,j>1}(-1)^{(j-1)i}(-1)^{n-1}l_jl_il_1+ (-1)^{n-1}l_1l_nl_1$$

$$=\sum_{i,j>1}(-1)^{(j-1)i}(-1)^{n-1}l_jl_il_1+
(-1)^{n-1}(-\sum_{i,j>1}(-1)^{(j-1)i}
l_jl_il_1)=0$$ and so again, we have the existence of $l_n$ together
with the appropriate sh Lie relations.}

${\bf Remarks:}$
(i) It may be the case in practice that the complex $X_*$ is truncated
at height $n$, i.e. we have that $X_*$ is not a resolution but rather
may have non-trivial homology in degree $n$ as well as in degree
$0$. In such a case, our construction of the maps $l_k$ may be
terminated by degree $n$ obstructions. More precisely, we have that
the vanishing of $H_kX_*$ for $k$ different from 0 and $n$ will then
only guarantee the existence of the requisite maps $l_k:(\bigotimes^k
X_*)_p \longrightarrow X_{p+k-2}$ for $p+k-2\leq n$. 

(ii) If property $(i)$ holds with zero on the right hand side, 
i.e. so that $l_2$ vanishes if one of the $x_i$'s is in ${\cal B}_0$, 
then $l_2$ can be extended trivially (to be a chain map) as zero on 
$(\bigotimes^2 X_*)_p$ for $p>0$. It is easy to see that in the
recursive construction, one can choose similarly trivial extensions of
the maps $l_k$ for $k>2$ to all of the resolution complex, i.e. they
are defined to vanish whenever one of their arguments 
belongs to ${\cal B}_0$
or $X_p$ for  $p>0$. Hence they induce well-defined maps $\hat l_k$ on
$\bigotimes^k{\cal F}$. With these choices, each of the defining
equations of the sh Lie algebra on $X_*$ involves only two terms,
namely $l_1l_k+l_{k-1}l_2=0$ for $k\geq 3$. 

(iii) If $H_kX_* = 0$ for $0<k<n$ and property (i) holds with zero on
the right hand side, we have defined a map on $\bigotimes^{n+2}{\cal
F}$ which may be non-zero. If so, the only non-trivial defining
equation of the induced sh Lie algebra on $\cal F$ 
reduces to $\hat l_{n+2}\hat l_2=0$. For example, if $n=1$, 
$\Sigma \hat l_3(\hat l_2(x_i,x_j),x_k,x_m) = 0$ where the
sum is over all permutations of (1234) such that $i<j$ and $ k < m$.

In section 3, we apply this construction in the context of Poisson
brackets in field theory. 

\subsection{Generalization to the graded case\label{gen}}

The above construction of an $L_\infty$-structure on the resolution of
a Lie algebra can be extended in a straightforward way to the graded
case, i.e. when the Lie algebra is graded (either by ${\bf Z}$ or by
${\bf Z/2}$, the super case) and the bracket is of a fixed degree,
even or odd, satisfying the appropriate graded version of
skew-symmetry and the Jacobi identity. We will refer to all of these
possibilities as graded Lie algebras although the older mathematical
literature uses that term only for the case of a degree 0 bracket. In
these situations, the resolution $X_*$ is bigraded and the inductive
steps proceed with respect to the resolution degree (see
\cite{jim:hrcpa,lars} for carefully worked out examples).  

The graded case occurs in the
Batalin-Fradkin-Vilkovisky approach to constrained Hamiltonian field
theories \cite{FV,BV1,FF,ht} where their bracket is of degree $0$ and in
the Batalin-Vilkovisky anti-field formalism for mechanical systems or
field theories \cite{bv:anti,bv:antired,ht} with their anti-bracket of
degree 1. 

In all these cases, one need only take care of the signs by the usual
rule: when interchanging two things (operators, fields, ghosts, etc.),
be sure to include the sign of the interchange.

\section{Local field theory with a Poisson bracket}

We first review the result that the cohomological resolution of local
functionals is provided by the horizontal complex. Then, we give the
definition of a Poisson bracket for local functionals. The existence
of such a Poisson bracket will assure us that the conditions of the
previous section hold. Hence, we show that to the Poisson bracket for
local functionals corresponds an sh Lie algebra on the graded
differential algebra of horizontal forms. 

\subsection{The horizontal complex as a resolution for local
  functionals.} 

In this subsection, we introduce some basic elements from jet-bundles
and the variational bicomplex relevant for our purpose. More details
and references to the original literature can be found in
\cite{Olver,saunders,dickey,ian}. For the most part, we will follow
the definitions and the notations of \cite{Olver}. Although much of
what we do is valid for general vector bundles, we will not be
concerned with global properties. We will use local coordinates most
of the time, though we will set things up initially in the global
setting.  

Let $M$ be an $n$-dimensional manifold and $\pi:E\rightarrow M$ a vector
bundle of fiber dimension $k$ over $M$. Let $J^{\infty}E$ denote the
infinite jet bundle of $E$ over $M$ with $\pi^{\infty}_E:J^{\infty}E
\rightarrow E$ and $\pi^{\infty}_M:J^{\infty}E \rightarrow M$ the
canonical projections. The vector space of smooth sections of $E$ with
compact support will be denoted $\Gamma E$. For each (local) section
$\phi$ of $E$, let $j^{\infty}\phi$ denote the induced (local) section
of the infinite jet bundle $J^{\infty}E$.

The restriction of the infinite jet bundle over  an appropriate open 
$U\subset M$ is trivial with fibre an infinite dimensional vector space
$V^\infty$.  The bundle
\begin{eqnarray}
\pi^\infty : J^\infty E_U=U\times V^\infty \rightarrow U 
\end{eqnarray} 
then has induced coordinates given by
\begin{eqnarray}
(x^i,u^a,u^a_i,u^a_{i_1i_2},\dots,).
\end{eqnarray}
We use multi-index notation and the summation convention throughout the
paper. If $j^{\infty}\phi$ is the section of $J^{\infty}E$ induced by
a section $\phi$ of the bundle $E$, then $u^a\circ
j^{\infty}\phi=u^a\circ \phi$ and $$u^a_I\circ j^{\infty}\phi=
(\partial_{i_1}\partial_{i_2}...\partial_{i_r})(u^a\circ
j^{\infty}\phi)$$ where $r$ is the order of the symmetric multi-index
$I=\{i_1,i_2,...,i_r\}$,with the convention that, for $r=0$, there are
no derivatives. 

The de Rham complex of differential forms $\Omega^*(J^{\infty}E,d)$ on
$J^{\infty}E$ possesses a differential ideal, the ideal ${ C}$ of
contact forms $\theta$ which satisfy $(j^{\infty}\phi)^* \theta=0$ for
all sections $\phi$ with compact support. This ideal is generated by
the contact one-forms, which in local coordinates assume the form
$\theta^a_J=du^a_J-u^a_{iJ}dx^i$. Contact one-forms of order $0$
satisfy $(j^{1}\phi)^*(\theta)=0$. In local coordinates, contact forms
of order zero assume the form $\theta^a= du^a-u^a_idx^i$.

Remarkably, using the contact forms, we see that the complex
$\Omega^*(J^{\infty}E,d)$ splits as a bicomplex (though the finite
level complexes $\Omega^*(J^pE)$ do not). The bigrading is described
by writing a differential $p$-form $\alpha$ as an element of
$\Omega^{r,s}(J^{\infty}E)$, with $p=r+s$ when
$\alpha=\alpha_{IA}^{\bf J}(\theta^A_{\bf J}\wedge dx^I)$ where
\begin{eqnarray} dx^I=dx^{i_1}\wedge...\wedge dx^{i_r} \quad {\rm and}
  \quad \theta^A_{\bf J}=\theta^{a_1}_{J_1}\wedge...\wedge
  \theta^{a_s}_{J_s}. \end{eqnarray} 
We intend to restrict the complex $\Omega^*$ by requiring that the
functions $\alpha_{IA}^{\bf J}$ be local functions in the following
sense. 

\begin{definition}
A {\bf local function} on $J^{\infty}E$ is the pullback of a smooth
function on some finite jet bundle $J^pE$, i.e. a composite $J^\infty
E \to J^pE \to M$. In local coordinates, a local function
$L(x,u^{(p)})$ is a smooth function in the coordinates $x^i$ and the
coordinates $u^a_I$, where the order $|I| = r$ of the multi-index $I$
is less than or equal to some integer $p$. 

The {\bf space of local functions} will be denoted $\CJ$, while the
subspace consisting of functions $(\pi^{\infty}_M)^*f$ for $f\in
C^{\infty}M$ is denoted by $Loc_M.$
\end{definition}

Henceforth, the coefficients of all differential forms in the complex
\newline 
$\Omega^*(J^{\infty}E,d)$ are required to be local functions, i.e., for
each such form $\alpha$ there exists a positive integer $p$ such that
$\alpha$ is the pullback of a form of $\Omega^*(J^pE,d)$ under the
canonical projection of $J^{\infty}E$ onto $J^pE$. In this context,
the horizontal differential is obtained by noting that $d\alpha$ is in
$\Omega^{r+1,s}\oplus \Omega^{r,s+1}$ and then denoting the two pieces
by, respectively, $d_H \alpha$ and $d_V\alpha$. One can then write 
\begin{eqnarray}
d_H \alpha=(-1)^s\{D_i \alpha_{IA}^{\bf J}\theta^A_{\bf J}\wedge
dx^i\wedge dx^I +\alpha_{IA}^{\bf J}\theta^A_{{\bf J}i}\wedge
dx^i\wedge dx^I \} , 
\end{eqnarray}
where
$$\theta^A_{{\bf J}i}=\sum_{r=1}^s
(\theta^{a_1}_{J_1}\wedge...\theta^{a_r}_{J_ri}...\wedge
\theta^{a_s}_{J_s})$$ and where
\begin{eqnarray}
D_i=\frac {\partial}{\partial x^i}+u^a_{iJ}\frac {\partial}{\partial
u^a_J} 
\end{eqnarray}
is the total differential operator acting on local functions.

We will work primarily with the $d_H$ subcomplex, the algebra of
horizontal forms $\Omega^{*,0}$, which is the exterior algebra in the
$dx^i$ with coefficients that are local functions. In this case we
often use Olver's notation $D$ for the horizontal differential
$d_H=dx^iD_i$ where $D_i$ is defined above. In addition to this
notation, we also utilize the operation $\lh$ which is defined as
follows. Given any differential r-form $\alpha$ on a manifold $N$ and
a vector field $X$ on $N$, $X\lh \alpha$ denotes that (r-1)-form whose
value at any $x\in N$ and $(v_1,...,v_{r-1})\in (T_xN\times \cdots
\times T_xN)$ is $\alpha_x(X_x,v_1,...,v_{r-1}).$ We will sometimes
use the notation $X(\alpha)$ in place of $X\lh\alpha$. 

Let $\nu$ denote a fixed volume form on  $M$ and let $\nu$ also denote
its pullback $(\pi^{\infty}_E)^*(\nu)$ to $J^{\infty}E$ so that $\nu$
may be regarded either as a top form on $M$ or as defining elements
$P\nu$ of $\Omega^{n,0}E$ for each $P\in C^{\infty}(J^{\infty}E)$. We
will almost invariably assume that $\nu = d^nx =
dx^1\wedge\cdots\wedge dx^n$. 

It is useful to observe that for $R^i\in Loc(E)$ and 
\begin{eqnarray}
\alpha=(-1)^{i-1}R^i(\frac{\partial}{\partial x^i} \lh d^nx),
\end{eqnarray} then
\begin{eqnarray}
d_H \alpha= (-1)^{i-1} D_jR^i[dx^j\wedge (\frac {\partial}{\partial
x^i} \lh  d^nx)]=D_jR^jd^nx. 
\end{eqnarray}
Hence, a total divergence $D_jR^j$ may be represented (up to the
insertion of a volume $d^nx$) as the horizontal differential of an
element of $\Omega^{n-1,0}(J^{\infty}E)$. It is easy to see that
the converse is true so, that, in local coordinates, one has that
total divergences are in one-to-one correspondence with $D$-exact
n-forms. 

\begin{definition}
A {\bf local functional}
\begin{eqnarray}
{\cal L}[\phi]=\int_M L(x,\phi^{(p)}(x)) dvol_M = \int_M (j^\infty
\phi)^*  L(x,u^{(p)}) dvol_M 
\end{eqnarray}
is the integral over $M$ of a local function evaluated for sections
$\phi$ of $E$ of compact support.

The {\bf space of local functionals} ${\cal F}$ is the vector space of
equivalence classes of local functionals, where two local functionals
are  equivalent if they agree for all sections of compact support.
\end{definition}

If one does not want to restrict oneself to the case where the base
space is a subset of ${\bf R}^n$, one has to take the transformation
properties of the integrands under coordinate transformations into
account and one has to integrate a horizontal $n$-form rather than a
multiple of $dx^n$ by an element of $Loc(E)$. 

\begin{lemma}\label{l4}
The vector space of local functionals ${\cal F}$ is isomorphic to the
cohomology group $H^n(\Omega^{*,0},D)$.
\end{lemma}
\proof{ Recall that one has a natural mapping $\hat{\eta}$ from
  $\Omega^{n,0}(J^{\infty}E)$ onto ${\cal F}$ defined by
\begin{eqnarray}
\hat{\eta}(P\nu)(\phi)=\int_M (j^{\infty}\phi)^*(P)\nu \quad\quad
\forall \phi\in \Gamma E.
\end{eqnarray}
It is well-known (see e.g. \cite{Olver}) that
$\hat{\eta}(P\nu)[\phi]=0$ for all $\phi$ of compact support if and
only if in coordinates $P$ may be represented as a divergence, i.e.,
iff $P=D_iR^i$ for some set of local functions $\{R^i\}$. Hence,
$\hat{\eta}(P\nu)=0$ if and only if there exists a form $\beta\in
\Omega^{n-1,0}$ such that the horizontal differential $d_H$ maps
$\beta$ to $P\nu$. So the kernel of $\hat{\eta}$ is precisely
$d_H\Omega^{n-1,0}$ and $\hat{\eta}$ induces an isomorphism from
$H^n(d_H)=\Omega^{n,0}/ d_H\Omega^{n-1,0}$ onto the space ${\cal F}$
of local functionals. }

For later use, we also note that the kernel of $\hat{\eta}$ coincides
with the kernel of the Euler-Lagrange operator: for $1\leq a\leq m$,
let $E_a$ denote the $a$-th component of the Euler-Lagrange operator
defined for $P\in Loc(E)$ by 
\begin{eqnarray}
E_a(P)=\frac{\partial P}{\partial u^a}-\partial_i\frac{\partial
P}{\partial u^a_i}+\partial_i\partial_j\frac{\partial P}{\partial
u^a_{ij}}-...=(-D)_I(\frac{\partial P}{\partial u^a_I}).
\end{eqnarray}
The set of components $\{E_a(P)\}$ are in fact the components of a
covector density with respect to the generating set $\{\theta^a\}$ for
${ C}_0$, the ideal generated by the contact one-forms of order
zero. Consequently, the Euler operator 
\begin{eqnarray} E(Pd^nx)=E_a(P)(\theta^a\wedge d^nx), 
\end{eqnarray}
for $\{\theta^a\}$ a basis of $ { C}_0$, gives a well-defined element of
$\Omega^{n,1}$. We have $E(P\nu)=0$ iff $P\nu=d_H\beta$. For
convenience we will also extend the operator $E$ to map local
functions to $\Omega^{n,1}$, so that $E(P)$ is defined to be
$E(Pd^nx)$ for each $P\in Loc(E)$.

In section 2, we have assumed that we have a resolution of 

\noindent ${\cal F}\simeq  H^n(\Omega^{*,0},d_H)$. 
In the case where $M$ is contractible, such a
resolution  necessarily exists and is provided by the following
(exact) extension of the horizontal complex
$\Omega^{*,0}(Loc(E),d_H)$:\\
$\begin{array}{ccccccccccc}\\& & & d_H & & d_H & & d_H & & d_H&\\ {\bf
    R}& \longrightarrow &\Omega^{0,0}&\longrightarrow &\dots &
\longrightarrow &
\Omega^{n-2,0} &\stackrel {\textstyle}{\longrightarrow}
&\Omega^{n-1,0}&\stackrel {\textstyle}{\longrightarrow}
&\Omega^{n,0}\\ \downarrow\eta & &\downarrow\eta & & & & 
\downarrow\eta & & \downarrow\eta & & \downarrow\eta\\ 0 &
\longrightarrow & 0 &\longrightarrow &\dots & \longrightarrow
&0&\longrightarrow & 0 &\longrightarrow & H_0={\cal F} 
\end{array}$
\vskip1ex

Alternatively, we can achieve a resolution by taking out the
constants: the space $X_*$ is given by $X_i=\Omega^{n-i,0}$, for
$0\leq i < n$, $X_n=\Omega^{0,0}/{\bf R}$, and $X_i=0$, for
$i>n+1$. Either way, we have a resolution of ${\cal F}$ and can
proceed with the construction of an sh Lie structure. (For general
vector bundles $E$, the assumption that such a resolution  exists
imposes topological restrictions on $E$ which can be shown to depend
only on topological properties of $M$ \cite{ian}.) 

\subsection{Poisson brackets for local functionals}

To begin to apply the results of section 2, we must have a bilinear
skew-symmetric mapping $\tilde l_2$ from $\Omega^{n,0}\times
\Omega^{n,0}$ to $\Omega^{n,0}$ such that: 

(i) $\tilde l_2(\alpha,d_H\beta)$ belongs to $d_H\Omega^{n-1,0}$ for 
all $\alpha\in\Omega^{n,0}$ and $\beta\in \Omega^{n,0}$, and

(ii) $\sum_{\sigma\in unsh(2,1)}(-1)^{|\sigma|}\tilde l_2(\tilde
l_2(\alpha_{\sigma(1)},\alpha_{\sigma(2)}),\alpha_{\sigma(3)})$
belongs to $d_H\Omega^{n-1,0}$ for all $\alpha_1,\alpha_2,\alpha_3\in
\Omega^{n,0}$. 

To introduce a candidate $\tilde l_2$, we define additional
concepts. We  say that $X$ is a {\bf generalized vector field over}
$M$ iff $X$ is a mapping which factors through the differential of the
projection of $J^\infty E$ to $J^rE$ for some non-negative integer r
and which assigns to each $w\in J^\infty E$ a tangent vector to $M$ at
$\pi^{\infty}_M(w)$. Similarly $Y$ is a {\bf generalized vector field
over} $E$ iff $Y$ also factors through $J^rE$ for some $r$ and assigns
to each $w\in J^\infty E$ a vector tangent to $E$ at
$\pi^{\infty}_E(w)$. In local coordinates one has 
\begin{eqnarray}
X=X^i\partial/\partial x^i \quad \quad Y= Y^j\partial/\partial x^j
+Y^a\partial/\partial u^a 
\end{eqnarray} where $X^i,Y^j,Y^a\in Loc(E)$. A generalized vector
field $Q$ on $E$ is called an evolutionary vector field iff
$(d\pi)(Q_w)=0$ for all $w\in J^{\infty}E$. In adapted coordinates, an
evolutionary vector field assumes the form $Q=Q^a(w) \partial/\partial
u^a$. 

Given a generalized vector field $X$ on $M$, there exists a unique
vector field denoted Tot(X) such that $(d\pi^{\infty}_M)(Tot(X))=X$
and $\theta(Tot(X))=0$ for every contact one-form $\theta$. In the
special case that $X=X^i\partial/\partial x^i $, it is easy to show
that $Tot(X)=X^iD_i$. We say that $Z$ is a first order total differential
operator iff there exists a generalized vector field $X$ on $M$ such
that $Z=Tot(X)$. More generally, a total differential operator $Z$ is
by definition the sum of a finite number of finite order operators
$Z_{\alpha}$ for which there exists functions $Z^J_{\alpha} \in
Loc(E)$ and first order total differential operators $W_1,W_2,...W_p$
on $M$ such that 
\begin{eqnarray}
Z_{\alpha}=Z^J_{\alpha}(W_{j_1}\circ W_{j_2}\circ ...\circ W{j_p})
\end{eqnarray}
where $J=\{j_1,j_2,...,j_p\}$ is a fixed set of multi-indices
depending on  $\alpha$ ($p = 0$ is allowed).

In particular, in adapted coordinates, a total differential operator
assumes the form $Z=Z^JD_J$, where $Z^J\in Loc(E)$ for each
multi-index $J$, and the sum over the multi-index $J$ is restricted to
a finite number of terms. 

In an analogous manner, for every evolutionary vector field $Q$ on
$E$, there exists its {\bf prolongation}, the unique vector field
denoted $pr(Q)$ on $J^{\infty}E$ such that
$(d\pi^{\infty}_E)(pr(Q))=Q$ and ${\cal L}_{pr(Q)}({ C})\subseteq {
C}$, where ${\cal L}_{pr(Q)}$ denotes the Lie 
derivative operator with respect to the vector field $pr(Q)$ and ${
C}$ is the ideal of contact forms on $J^{\infty}E$.
In local adapted coordinates, the prolongation of an evolutionary
vector field $Q=Q^a\partial/\partial u^a$ assumes the form
$pr(Q)=(D_JQ^a)\partial/\partial u^a_J$. The set of all total
differential operators will be denoted by $TDO(E)$ and the set
of all evolutionary vector fields by $Ev(E)$. Both $TDO(E)$ and
$Ev(E)$ are left $Loc(E)$ modules. 

One may define a new total differential operator $Z^+$ called the {\bf
adjoint} of $Z$ by 
\begin{eqnarray}
\int_M (j^{\infty}\phi)^*[SZ(T)]\nu=\int_M
(j^{\infty}\phi)^*[Z^+(S)T]\nu 
\end{eqnarray}
for all sections $\phi\in \Gamma E$ and all $S,T\in Loc(E)$. It
follows that 
\begin{eqnarray}
[SZ(T)]\nu=[Z^+(S)T]\nu + d_H\zeta\label{x} 
\end{eqnarray} for some
$\zeta\in \Omega^{n-1,0}(E)$. If $Z=Z^JD_J$ in local coordinates, then
$Z^+(S)=(-D)_J(Z^JS)$. This follows from an integration by parts in
(\ref{x}) and the fact that (\ref{x}) must hold for all $T$ (see e.g.
\cite{Olver} corollary 5.52). 

Assume that $\omega$ is a mapping from ${ C}_0 \times { C}_0$ to $TDO(E)$
which is a module homomorphism in each variable separately. The
adjoint of $\omega$ denoted $\omega^+$ is the mapping from ${ C}_0
\times { C}_0$ to $TDO(E)$ defined by
$\omega^+(\theta_1,\theta_2)=\omega(\theta_2,\theta_1)^+$. In
particular  $\omega$ is skew-adjoint iff $\omega^+=-\omega$.

Using the module basis $\{\theta^a\}$ for ${ C}_0$ , we define the
total differential operators
$\omega^{ab}=\omega(\theta^a,\theta^b)$. From these operators, we
construct a bracket on the set of local functionals 
\cite{gel1,gel2,gel3,gel4,gel5} (see e.g. \cite{Olver,dickey} for 
reviews) by  
\begin{eqnarray} \{{\cal P},{\cal Q}\}=\int_M
\omega(\theta^a,\theta^b)(E_b(P))E_a(Q) d^nx, 
\end{eqnarray} where ${\cal P}=Pd^nx$ and ${\cal Q}=Qd^nx$ for local
functions $P$ and $Q$. As in other formulas of this type, it is
understood that the local functional $\{{\cal P},{\cal Q}\}$ is to be
evaluated at a section $\phi$ of the bundle $E\rightarrow M$ and that
the integrand is pulled back to $M$ via $j^{\infty}\phi$ before being
integrated. 

We find it useful to introduce the condensed notation
$\omega(E(P))E(Q)= \omega(\theta^a,\theta^b)(E_b(P))E_a(Q)$ throughout
the remainder of the paper. 
In order to express $\omega(E(P))E(Q)$ in a coordinate invariant
notation, note that 

\noindent $pr(\frac{\partial}{\partial u^a})\lh
E(L)=E_a(L)d^nx$ for each local function$L$. Consequently, if $*$ is
the operator from $\Omega^{n,0}E$ to $Loc(E)$ defined by $*(P\nu)=P$,
then 
\begin{eqnarray} \omega(E(P))E(Q)=\omega(\theta^a,\theta^b)(
*[pr(\frac{\partial}{\partial u^b})\lh
E(P)])(*[pr(\frac{\partial}{\partial u^a})\lh E(Q)]). 
\end{eqnarray}
If coordinates on $M$ are chosen such that $\nu= d^nx$, then it
follows that 
\begin{eqnarray}
\{{\cal P},{\cal Q}\}=\int_M \omega(\theta^a,\theta^b)( *[pr(Y_b)\lh 
E(P)])(*[pr(Y_a)\lh E(Q)])\nu, 
\end{eqnarray} 
where $\{Y_a\}$ and $\{\theta^b\}$ are required to be local bases of
$Ev(E)$ and $C_0$, respectively, such that
$\theta^b(Y_a)=\delta^b_a$. It is easy to show that the integral is
independent of the choices of bases and consequently, one has a
coordinate-invariant description of the bracket for local
functionals. 

\subsection{Associated sh Lie algebra on the horizontal complex}

This bracket for functionals provides us with some insight as to how
$\tilde l_2$ may be defined; namely for $\alpha_1=P_1\nu$ and
$\alpha_2=P_2\nu\in \Omega^{n,0}$, define $\tilde
l_2(\alpha_1,\alpha_2)$ to be the skew-symmetrization of the integrand
of $\{{\cal P}_1,{\cal P}_2\}$~: \begin{eqnarray} \tilde
l_2(\alpha_1,\alpha_2)
=\frac{1}{2}[\omega(E(P_1))E(P_2)-\omega(E(P_2))E(P_1)]\nu.\label{z}
\end{eqnarray}
By construction, $\tilde l_2$ is skew-symmetric and, from
$E(d_H\beta)=0$ for $\beta\in\Omega^{n-1,0}$, it follows that $\tilde
l_2(\alpha,d_H\beta)=0$. Thus a strong form of property (i) required
above for $\tilde l_2$ holds. 

The symmetry properties of $\omega$ may be used to simplify the
equation for $\tilde l_2(\alpha_1,\alpha_2)$. Skew-adjointness of
$\omega$ implies 
\begin{eqnarray}
\omega(E(P_1))E(P_2)\nu=-\omega(E(P_2))E(P_1)\nu+d_H\gamma \label{w}
\end{eqnarray}
for some $\gamma\in\Omega^{n-1,0}$, which depends on $\alpha_1=P_1\nu$
and $\alpha_2=P_2\nu$. In fact, since $E(d_H\beta)=0$, the element
$\gamma$ depends only on the cohomology classes ${\cal P}_1,{\cal
P}_2$ of $\alpha_1$ and $\alpha_2$. A specific formula for $\gamma$
can be given by straightforward integrations by parts.

Hence, from (\ref{z}) and (\ref{w}), we get 
\begin{eqnarray} \tilde
l_2(\alpha_1,\alpha_2)=\omega(E(P_1))E(P_2)\nu-\frac{1}{2}d_H 
\gamma({\cal P}_1,{\cal P}_2).
\end{eqnarray}
Furthermore, since $\int_M(j^{\infty}\phi)^*d_H\gamma=0$ for all
$\phi\in \Gamma  E$, we see that
\begin{eqnarray}
\{{\cal P}_1,{\cal
P}_2\}(\phi)=\int_M(j^{\infty}\phi)^*[\omega(E(P_1))E(P_2)]\nu=
\int_M(j^{\infty}\phi)^*\tilde l_2(\alpha_1,\alpha_2). 
\end{eqnarray}

In order to explain the conditions necessary for $\tilde l_2$ to
satisfy the required ``Jacobi'' condition, we formulate the problem in
terms of ``Hamiltonian'' vector fields (see e.g. \cite{Olver} chapter
7.1 or \cite{dickey} chapter 2.5) and their corresponding Lie brackets.

Given a local function $Q$, one defines an evolutionary vector field
$v_{\omega EQ}$ by 
\begin{eqnarray}
v_{\omega EQ}= \omega^{ab}(E_b(Q))\partial/\partial u^a=
\omega(\theta^a,\theta^b)(*[pr(\partial/\partial u^b)\lh E(Q)])
\partial/\partial u^a.
\end{eqnarray}
Again, the vector field $v_{\omega EQ}$ depends only on the functional
${\cal Q}$ and not on which representative $Q$ one chooses in the
cohomology class ${\cal Q}\sim Q\nu+d_H\Omega^{n-1,0}$. Thus, for a
given functional ${\cal Q}$, let $\hat{v}_{{\cal Q}}=v_{\omega EQ}$ for
any representative $Q$.

Since $\{{\cal P}_1,{\cal P}_2\}=\int_M \tilde
l_2(\alpha_1,\alpha_2)$, we see that
\begin{eqnarray}
\hat{v}_{\{{\cal P}_1,{\cal P}_2\}}=
v_{\omega E(\tilde l_2(\alpha_1,\alpha_2))}.\label{u}
\end{eqnarray}

Note also that
\begin{eqnarray}
\omega (E(P_1))E(P_2)=\omega^{ab}(E_b(P_1))E_a(P_2)=
pr[\omega^{ab}E_b(P_1)\partial/\partial u^a]\lh E(P_2)\nonumber\\
=pr(v_{\omega E(P_1)})\lh E(P_2)=pr(\hat{v}_{{\cal P}_1})\lh E(P_2).
\end{eqnarray}
Moreover, integration by parts allows us to show that 
\begin{eqnarray}
pr(Q)(P)\nu=pr(Q)\lh d(P\nu)=pr(Q)\lh E(P\nu) +d_H(pr(Q)\lh \sigma),
\end{eqnarray}
for arbitrary evolutionary vector fields $Q$ and local functions $P$,
and for some form $\sigma\in\Omega^{n-1,0}$ depending on $P$. For
every such $Q$, let $I_Q$ denote a mapping from $\Omega^{n,0}$ to
$\Omega^{n-1,0}$ such that 
\begin{eqnarray}
pr(Q)(P)\nu=pr(Q)\lh E(P) +d_H(I_Q(P\nu)) 
\end{eqnarray} 
for all $P\nu\in\Omega^{n,0}$. Explicit coordinate expressions for
$I_Q$ can be found in \cite{Olver} chapter 5.4 or in \cite{dickey}
chapter 17.5 .

It follows from the identities (32), (36) and (38) that 
\begin{eqnarray} 
\tilde l_2(\alpha_1,\alpha_2)=pr(\hat{v}_{{\cal
P}_1})(P_2)\nu-d_HI_{\hat{v}_{{\cal P}_1}}(P_2\nu)-\frac{1}{2}d_H
\gamma({\cal P}_1,{\cal P}_2)).\label{t} 
\end{eqnarray}
Thus, for $\alpha_1,\alpha_2,\alpha_3\in \Omega^{n,0}$, we see that
\begin{eqnarray}
\tilde l_2(\tilde l_2(\alpha_1,\alpha_2),\alpha_3)= 
-\tilde l_2(\alpha_3,\tilde l_2(\alpha_1,\alpha_2))=\hskip15ex \ \ \ \
\ \nonumber\\   \hskip15ex =-\tilde l_2(\alpha_3,pr(\hat{v}_{{\cal
P}_1})(P_2)\nu-d_H(\cdot))= -\tilde l_2(\alpha_3,pr(\hat{v}_{{\cal
P}_1})(P_2)\nu) 
\end{eqnarray} 
and
\begin{eqnarray}
\tilde l_2(\tilde l_2(\alpha_1,\alpha_2),\alpha_3)= -pr(\hat{v}_{{\cal
P}_3})(pr(\hat{v}_{{\cal P}_1})(P_2))\nu+d_H\zeta,\label{v} 
\end{eqnarray} 
where $\zeta$ is given by
\begin{eqnarray}
\zeta( {\cal P}_1, P_2,{\cal P}_3)
=I_{\hat{v}_{{\cal P}_3}}(pr(\hat{v}_{{\cal
P}_1})(P_2)\nu)+\frac{1}{2}\gamma({\cal
P}_3,\{{\cal P}_1,{\cal P}_2\}).
\end{eqnarray}
Rewriting the left hand side of the Jacobi identity in Leibnitz form
and using (\ref{u}), (\ref{t}) and (\ref{v}), we find 
\begin{eqnarray} 
\sum_{\sigma\in unsh(2,1)}(-1)^{|\sigma|}\tilde l_2(\tilde
l_2(\alpha_{\sigma(1)},\alpha_{\sigma(2)}),\alpha_{\sigma(3)})
=\nonumber\\  = -\tilde l_2(\alpha_{3},\tilde
l_2(\alpha_{1},\alpha_{2})) -\tilde l_2(\tilde
l_2(\alpha_{1},\alpha_{3}),\alpha_{2}) +\tilde l_2(\alpha_{1},\tilde
l_2(\alpha_{3},\alpha_{2})) =\nonumber\\ =\lbrack -pr(\hat{v}_{{\cal
P}_3})(pr(\hat{v}_{{\cal P}_1})(P_2))+pr(\hat{v}_{{\cal
P}_1})(pr(\hat{v}_{{\cal P}_3})(P_1)) \nonumber\\
-pr(\hat{v}_{ \{ {\cal P}_1,{\cal P}_3 \} })(P_2)\rbrack\nu+ d_H \eta,
\end{eqnarray}
with
\begin{eqnarray}
\eta({\cal P}_1,{\cal P}_2,{\cal P}_3)=
\zeta({\cal P}_1, P_2,{\cal P}_3)-
\zeta({\cal P}_3,P_2,{\cal P}_1)\nonumber\\+ I_{\hat{v}_{\{{\cal
P}_1,{\cal  P}_3\}}}(P_2\nu)+\frac{1}{2}d_H \gamma(\{{\cal P}_1,{\cal
P}_3\},{\cal P}_2). \label{boun}
\end{eqnarray}
Although $\zeta$ depends on the representative $P_2$ and not its cohomology 
class, $\eta$ depends only on the cohomology classes ${\cal P}_i$
because it is completely skew-symmetric.

It follows from this identity that if
\begin{eqnarray}
pr(\hat{v}_{\{{\cal P}_1,{\cal P}_2\}})=[pr(\hat{v}_{{\cal
P}_1}),pr(\hat{v}_{{\cal P}_2})]
\end{eqnarray}
for all ${\cal P}_1,{\cal P}_2$, then $\{\cdot,\cdot \}$ satisfies the
Jacobi condition. Under these conditions, the mapping $H: {\cal
F}\longrightarrow Ev(E)$ defined by $H({\cal P})=\hat v_{{\cal P}}$ is
said to be {\bf Hamiltonian}. Equivalent conditions on the mapping
$H$ alone for the bracket $\{\cdot,\cdot \}$ to be a Lie bracket can
be found in \cite{Olver,dickey}. The derivation given here allows us
to give, in local coordinates, an explicit form for the exact term
(\ref{boun}) violating the Jacobi identity.

If $H$ is Hamiltonian, the bracket $\tilde l_2$ satisfies
condition (ii) and the construction of section 2 applies. Because the
resolution stops with the horizontal zero-forms, we get a possibly
non-trivial $L(n+2)$ algebra on the horizontal complex. If we remove
the constants, we can then extend to a full $L_\infty -$algebra by
defining the further $l_i$ to be $0$. In addition, because property
(i) holds without any $l_1$ exact term on the right hand side, remark
(ii) at the end of section 2.1 applies, i.e., we need only two terms
in the defining equations of the sh Lie algebra and the maps $l_k$
induce well-defined higher order maps on the space of local
functionals. On the other hand, if we do not remove the constants, the
operation $l_{n+2}$ takes values in $X_n=\Omega^{0,0}= Loc(E)$
and induces a multi-bracket on $H^n(\Omega^{*,0},d_H)\simeq {\cal
F}$, the space of local functionals, with values in $H_n(X_*,l_1) =
H^0(\Omega^{*,0},d_H) \simeq H_{DR}(C^\infty(M))={\bf R}.$ 

We have thus proved the following main theorem. 

\begin{theorem} Suppose that the horizontal complex without the
  constants 

  \noindent $(\Omega^{*,0}/{\bf R},d_H)$ is a resolution of
  the space of local functionals ${\cal F}$ equipped with a Poisson
  bracket as above. If the mapping $H$ from ${\cal F}$ to
  evolutionary vector fields is Hamiltonian, then to the Lie algebra
  ${\cal F}$ equipped with the induced bracket $\hat l_2=\{\cdot,\cdot\}$, 
  there correspond maps $l_i:(\Omega^{*,0}/{\bf R})^{\otimes i}\to
\Omega^{*-i+2,0}/{\bf R}$ for
  $1\leq i \leq n+2$ satisfying the sh Lie identities  $$l_1 l_k +
  l_{k-1} l_2=0.$$ The corresponding map $\hat l_{n+2}$ 
on ${\cal F}^{\otimes n+2}$ with
  values in $H^0(\Omega^{*,0},d_H)={\bf R}$ satisfies $$\hat
  l_{n+2}\hat l_2=0. $$
\end{theorem}

Specific examples for $n=1$ are worked out in careful detail by Dickey
\cite {dickey:ex}.

\section{Conclusion}

The approach of Gel'fand, Dickey and Dorfman to functionals and 
Poisson brackets in field theory has the advantage of being completely
algebraic. In this paper, we have kept explicitly the boundary terms
violating the Jacobi identity for the bracket of functions, 
instead of throwing them away by going over to functionals at the
end of the computation. In this way, we can work consistently with
functions alone, at the price of deforming the Lie algebra into an sh Lie
algebra. Our hope is that this approach will be useful for
a completely algebraic study of deformations of Poisson brackets in
field theory.

\section*{Acknowledgments}
The authors want to thank I. Anderson, L.A. Dickey and M. Henneaux for useful
discussions. 

\ifx\undefined\bysame
\newcommand{\bysame}{\leavevmode\hbox to3em{\hrulefill}\,}
\fi

\end{document}